\documentclass[12pt]{iopart}
\usepackage{graphicx,bm}
\usepackage{txfonts}
\usepackage{dsfont}
\usepackage{color}
\usepackage{hyperref}

\usepackage[normalem]{ulem}

\usepackage{tikz}
\usepackage{lipsum}

\expandafter\let\csname equation*\endcsname\relax
\expandafter\let\csname endequation*\endcsname\relax

\usepackage{ae}
\usepackage{siunitx}
\usetikzlibrary{quantikz}
\usepackage{subcaption}
\usepackage{braket}
\usepackage{amsthm}
\usepackage{amssymb}
\begin{document}

\title{Deep-learned error mitigation via partially knitted circuits for the variational quantum eigensolver
}

\author{Simone Cantori$^{1,2}$, Andrea Mari$^1$, David Vitali$^{1,2}$, Sebastiano Pilati$^{1,2}$}
\address{$^1$ School of Science and Technology, Physics Division, University of Camerino, I-62032 Camerino (MC), Italy\\
$^2$ INFN-Sezione di Perugia, 06123 Perugia, Italy}

\vspace{10pt}
\begin{indented}
\item[]\today
\end{indented}

\begin{abstract}
The variational quantum eigensolver (VQE) is generally regarded as a promising quantum algorithm for near-term noisy quantum computers. 
However, when implemented with the deep circuits that are in principle required for achieving a satisfactory accuracy, the algorithm is strongly limited by noise.
Here, we show how to make VQE functional via a tailored error mitigation technique based on deep learning. Our method employs multilayer perceptrons trained on the fly to predict ideal expectation values from noisy outputs combined with circuit descriptors. Importantly, a circuit knitting technique with partial knitting is adopted to substantially reduce the classical computational cost of creating the training data. 
We also show that other popular general-purpose quantum error mitigation techniques do not reach comparable accuracies. 
Our findings highlight the power of deep-learned quantum error mitigation methods tailored to specific circuit families, and of the combined use of variational quantum algorithms and classical deep learning.
\end{abstract}


\section{Introduction}
The practical utilization of quantum computing is still challenged by the inherent noise and decoherence in the currently available devices~\cite{Preskill2018quantumcomputingin, RevModPhys.94.015004, StilckFrança2021, PhysRevX.10.041038}. In fact, noise is expected to be a key challenge also in the near and mid-term future \cite{zhang2022computing, preskill2025}. In this context, variational quantum algorithms~\cite{Cerezo_2021} have emerged as promising applications that might be less affected by noise. In particular, the variational quantum eigensolver (VQE) algorithm~\cite{Tilly_2022} has been proposed for estimating the ground-state energy of quantum systems. 
However, its performance is, in fact, affected by noise and it is unclear whether the high accuracies required for practical applications can be reached~\cite{noisy_vqe3,noisy_vqe4,noisy_vqe6,noisy_vqe8,noisy_vqe9}. Indeed, flexible variational ansatzes can be implemented using deep circuits, but the impact of noise rapidly increases with the circuit depth.

For the above reasons, intense research activity is being devoted to the development of error mitigation (EM) strategies~\cite{Cai_2023}. Popular approaches, such as zero-noise extrapolation (ZNE)~\cite{PhysRevLett.119.180509, PhysRevX.7.021050, Kandala_2019} and probabilistic error cancellation~\cite{PhysRevLett.119.180509, PhysRevX.8.031027, Zhang_2020, van_den_Berg_2023}, can reduce the impact of noise~\cite{noisy_vqe5}, but often come at the price of considerable sampling overhead and display limited effectiveness in the context of variational quantum algorithms~\cite{noisy_vqe7, Quek_2024, Takagi_2022}. Recently, EM techniques based on the integration of machine learning methods with noisy quantum data have been proposed~\cite{noisy_vqe1,Czarnik2021errormitigation,Cantori2024,PhysRevResearch.3.033098,Liao2025,Liao2024,Bennewitz2022, xu2025physicsinspiredmachinelearningquantum}. In particular, deep learning models are expected to succeed in learning the complex mapping between noisy measurements and their corresponding ideal values, effectively filtering out errors. However, a key challenge in using machine learning models is
the need for training data with exact expectation values, which demands excessive classical computational resources for generic large circuits. In previous studies, this problem was addressed by the training on easily simulable circuits, such as those representing product states or near-Clifford circuits~\cite{Czarnik2021errormitigation,PhysRevResearch.3.033098,Liao2025,PhysRevResearch.6.013223}. 
%
However, these circuits differ substantially from those employed in VQE, making them unsuitable for mitigating errors in this task.

\begin{figure*}
	\centering
	\includegraphics[width=\textwidth]{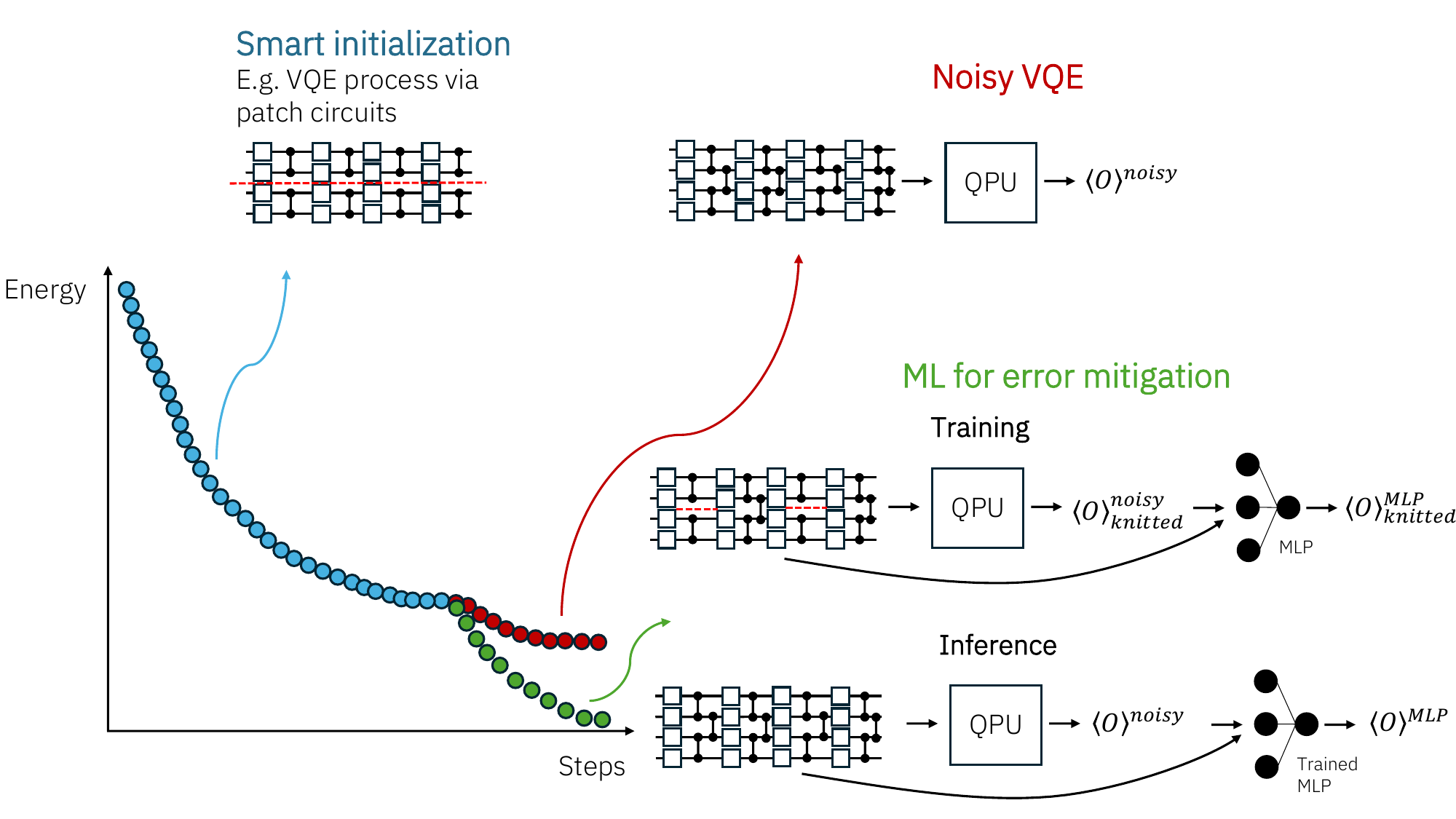}
	\caption{
    Schematic illustration of the VQE process enhanced by the deep learning-based error mitigation technique (DL-EM). First, a classical smart initialization of the variational parameters is applied performing a classical VQE optimization with simulable patch circuits. Secondly, before proceeding with the final VQE optimization, a neural network, which takes circuit descriptors and noisy expectation values as input data, is trained on partially-knitted quantum circuits having variational parameters close to the optimized configuration. In this training phase, the partially-knitted circuits are executed on the noisy quantum computer and on a classical simulator. 
    The VQE process is then completed using the trained neural network for evaluating error-mitigated expectation values. Note that, in the VQE steps, partial-knitting is not applied and the full variational circuit is executed on the quantum computer.
    } 
	\label{fig1}
\end{figure*}

In this article, we introduce a deep learning-based quantum error mitigation approach  (DL-EM) tailored to the VQE framework. DL-EM leverages artificial neural networks to predict noise-mitigated expectation values by processing hybrid quantum-classical input data composed of unmitigated expectation values (quantum data) and quantum circuit descriptors (classical data). 
Importantly, the network training is integrated within the VQE process, ensuring that the training circuits are representative of those required in successive stages of the variational optimization.
Furthermore, the computational cost for the generation of the training data is drastically reduced by incorporating the circuit knitting technique~\cite{Mitarai_2021}, appropriately adapted with partial knitting.
A schematic description of our approach is shown in Fig.~\ref{fig1}.

The testbed we consider is the one-dimensional (1D) quantum Ising model with random couplings.
Our simulations take into account realistic noise models, validated making comparisons against experiments executed on real IBM quantum devices. 
We find that while the unmitigated noisy VQE drastically fails to find the ground-state energy, the DL-EM method reaches an accuracy well below the 1\% threshold, with a further improvement as the amount of noise is reduced.
Instead, general-purpose EM methods, chiefly ZNE and corrections based on damping factors, do not provide comparable improvements.
Our combined strategy not only enhances EM but also paves the way for scalable implementations of VQE on current quantum hardware.
Our findings highlight the importance of EM methods tailored to specific applications and of the combined use of quantum computers with classical deep learning.

The remainder of this article is organized as follows: In Sec.~\ref {Sec2}, we describe the target Hamiltonian, the quantum circuits and the noise model, the (partial) circuit knitting technique employed in the generation of the training data, and the VQE process with the iterative DL-EM method. In Sec.~\ref{Sec3}, a set of numerical experiments is discussed, demonstrating that the DL-EM method substantially improves the accuracy of the noisy VQE algorithm. The performance of other popular general-purpose EM methods is also discussed. Our conclusions are reported in Sec.~\ref{Sec4}.

\section{Methods}\label{Sec2}
\subsection{Variational quantum eigensolver for the 1D quantum Ising model}\label{method1_sec}
\begin{figure*}
	\centering
	\includegraphics[width=\textwidth]{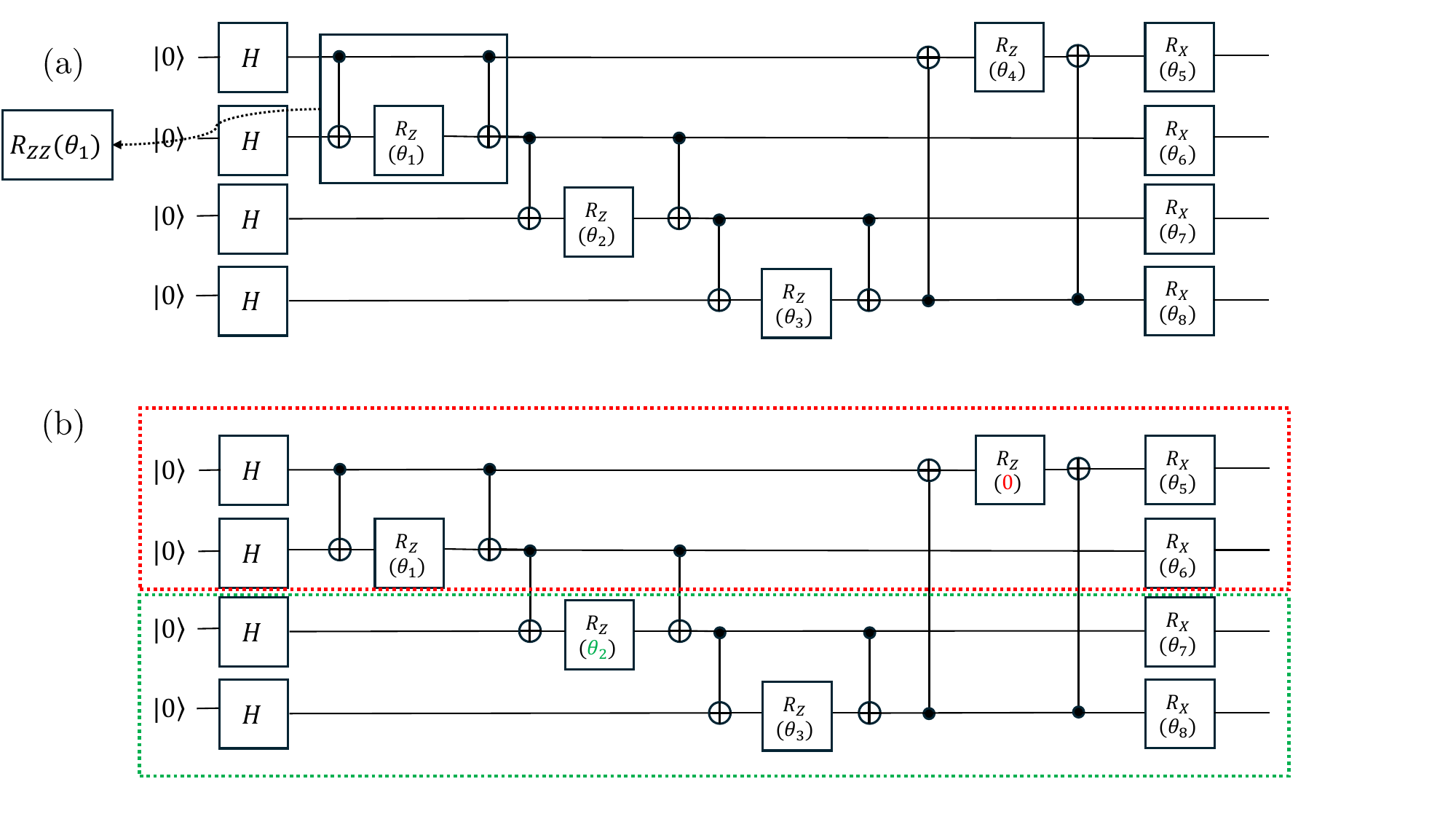}
	\caption{(a) Scheme of the first layer of the VQE ansatz. The same quantum gates, besides the initial superposition obtained with the Hadamard gates, are then repeated $P$ times with different parameters $\theta_i$. The decomposition of the $R_{ZZ}$ gate is also shown. (b) We divide the quantum circuit into 2 sub-circuits with the circuit knitting technique. Some of the parameters of the knitting gates are fixed to 0 (or $\pi$) effectively cutting the two partitions. Other parameters are left unchanged ($R_Z(\theta_2)$ in the figure), and a sampling overhead is needed to reconstruct the expectation values of the full circuit starting from the expectation values of the two sub-circuits, as explained in the main text.} 
	\label{knitting}
\end{figure*}

Our testbed for the VQE algorithm is the 1D quantum Ising model with nearest-neighbor random couplings. 
The Hamiltonian reads:
\begin{equation}\label{hamiltonian}
    H = -\sum_{i}J_{i}Z_iZ_{i+1} - h\sum_i X_i \, ,
\end{equation}
where $Z_i$ and $X_i$ are standard Pauli matrices, the couplings are sampled from a uniform distribution in the interval $J_{i} \in [-1,1]$, the transverse field is $h=0.5$, and periodic boundary conditions are adopted, i.e. $Z_{N+1}=Z_1$, with $N$ the number of spins.
VQE is a hybrid quantum-classical algorithm designed to approximate the ground-state energy of a quantum system~\cite{Tilly_2022}. Leveraging the variational principle, a parameterized quantum circuit (ansatz) $U(\boldsymbol{\theta})$ with parameters $\boldsymbol{\theta}$ is optimized to minimize the expectation value of the Hamiltonian. A quantum computer estimates the mean energy
\begin{equation}
\label{H1}
    E(\boldsymbol{\theta}) = \langle\psi(\boldsymbol{\theta}) |H|\psi(\boldsymbol{\theta})\rangle \, ,
\end{equation}
where $|\psi(\boldsymbol{\theta})\rangle = U(\boldsymbol{\theta})|0\rangle^{\otimes N}$ is the output state of the quantum circuit with input state $|0\rangle^{\otimes N}=|00...0\rangle$.
A classical optimizer updates the circuit parameters $\boldsymbol{\theta}$ iteratively, seeking for the minimum. We present results obtained using either gradient-based or gradient-free optimization strategies, specifically, employing the ADAM optimizer~\cite{adam} with the parameter-shift rule~\cite{PhysRevLett.118.150503, PhysRevA.98.032309}, and the COBYLA algorithm~\cite{2020SciPy-NMeth}, respectively.
We adopt the circuit architecture described in Ref.~\cite{10.21468/SciPostPhys.6.3.029}, which was originally designed for the pure ferromagnetic 1D quantum Ising model, corresponding to the Hamiltonian~\eqref{hamiltonian} with $J_i=1$ for $i=1,\dots,N$.
Since our model features random couplings, we extend the original circuit ansatz by assigning independent parameters to each gate, allowing different rotation angles for individual qubits across all circuit layers. The resulting ansatz is visualized in Fig.~\ref{knitting}a.

A key feature of our DL-EM method is that the neural networks are iteratively trained on the fly during the circuit optimization. The training is repeated when the target circuits differ substantially from those included in the training set.
Furthermore, the training circuits are created using the knitting technique, as further explained in Sec.~\ref{knitting_sec}. A partial knitting procedure allows us to find a balance between the classical computational cost and the similarity between training and target sets.
The VQE process starts with a smart parameter initialization, in which a variational optimization is performed with so-called  patch circuits~\cite{Arute2019}. These are formed by removing gates connecting two (or more) patches, allowing efficient classical execution.
Clearly, due to the disconnected structure, the patch circuits reach only an approximate representation of the ground state.
A schematic representation of the VQE process with our tailored EM method is shown in Fig.~\ref{fig1}. In summary, it involves the following steps:
\begin{enumerate}
  \item Smart parameter initialization with patch circuits;
  \item Training of the DL-EM neural network on partially knitted quantum circuits;
  \item Execution of VQE with DL-EM;
  \item Re-training of the DL-EM neural network when the circuit parameters deviate significantly from those used in the previous training phase;
  \item Repetition of steps 3 and 4 until convergence is reached.
\end{enumerate}

It is worth mentioning that other smart initialization criteria have been proposed in the literature~\cite{PhysRevApplied.22.054005,khan2023preoptimizingvariationalquantumeigensolvers}.
However, in our framework, patch circuits are particularly effective since the first steps of the parameter optimization involve small angles in the  $R_{ZZ}$ gates connecting the two patches. This feature reduces the computational cost of the circuit knitting technique employed in the training phase.

Even in the absence of noise, the accuracy reachable with VQE depends on the flexibility of the circuit ansatz. In fact, the required circuit depth was shown to scale with the size of the target system~\cite{10.21468/SciPostPhys.6.3.029, PhysRevResearch.5.013183}. 
For this reason, we adopt a relatively deep circuit featuring $P=8$ layers. 
Furthermore, in our simulations we adopt a realistic model of noise, as discussed in Sec.~\ref{noise_sec}. Therein, its effect is compared to the errors occurring in physical devices available from the IBM Quantum platform.
As detailed in Sec.~\ref{Sec3}, VQE with deep circuits is drastically affected by realistic noise, leading to poor performance. Some previous studies~\cite{noisy_vqe3,noisy_vqe5,Rosenberg_2022} reported relatively good accuracies of unmitigated noisy VQE, but considered shallow circuits suitable for small physical systems.

In Sec.~\ref{Sec3}, we also perform comparisons with other EM techniques. An important one is based on the observation that, if $\mathcal{O}$ is a traceless operator, its noisy expectation value under depolarizing noise is proportional to the exact expectation value $\langle\mathcal{O}\rangle_{\mathrm{noisy}}=D\langle\mathcal{O}\rangle$, where $D$ is the damping factor~\cite{PhysRevLett.127.270502,Rosenberg_2022}. By removing the single-qubit gates, one obtains a quantum circuit whose ideal result is known, because built using only CNOT gates. Hence, measuring the noisy expectation value allows determining $D$. The estimated damping factor can then be used to mitigate the expectation value of the original circuit by rescaling its noisy output. A similar approach can be used in combination with ZNE~\cite{PhysRevResearch.5.013183}. In our analysis, we consider the quantum circuit resulting from the last step of a VQE process, and we execute this error mitigation, dubbed damping-factor method (DM), both with and without ZNE. In both scenarios, the results are far less accurate than those achieved by DL-EM.
In Sec.~\ref{cost_sec}, we also discuss other machine learning-based EM methods and how they compare with our strategy.


\subsection{Noise model}\label{noise_sec}
\begin{figure*}
	\centering
	\includegraphics[width=\textwidth]{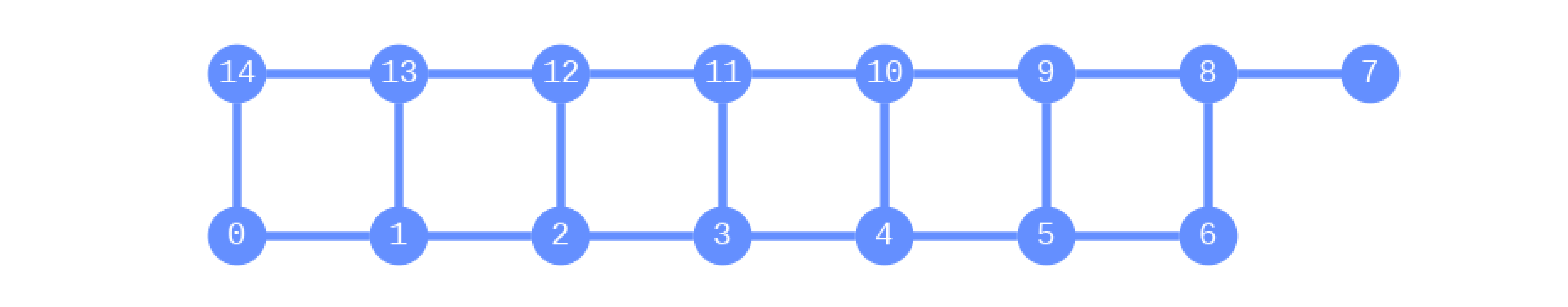}
	\caption{Layout of the IBM Melbourne quantum chip.} 
	\label{chip}
\end{figure*}

\begin{figure*}
	\centering
	\includegraphics[width=0.7\textwidth]{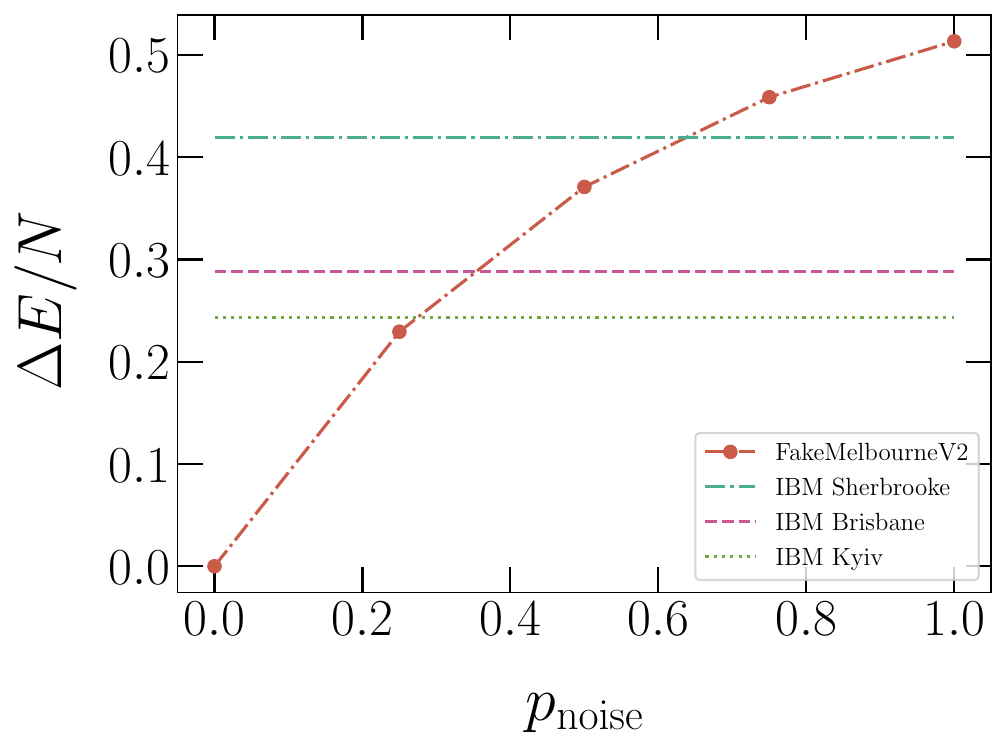}
	\caption{Energy discrepancies $\Delta E/N$ obtained by classically simulating the FakeMelbourneV2 model, as a function of the noise level $p_{\mathrm{noise}}$. These simulations are compared with actual experiments performed on the three quantum devices indicated in the legend, denoted by the horizontal lines. The three devices are accessed via the IBM quantum platform. The testbed is a quantum circuit representing the final step of a VQE process with $N=12$ qubits. The energy obtained at $p_{\mathrm{noise}}=0$ is the reference to calculate the discrepancy.} 
	\label{real}
\end{figure*}

The noisy expectation values are obtained by simulating a realistic quantum device using a noise model imported from Qiskit~\cite{qiskit2024}, namely, the FakeMelbourneV2 model. The chip layout corresponding to this device is shown in Fig.~\ref{chip}.
Notably, this layout allows preserving periodic boundary conditions across 1D quantum circuits of different sizes, up to $N=14$.
In our simulations, the noisy expectation values $\langle Z_iZ_{i+1}\rangle_{\mathrm{noisy}}$ and $\langle X_i\rangle_{\mathrm{noisy}}$ are obtained performing $S=10^6$ shots.
To thoroughly analyse the impact of noise, we customize the FakeMelbourneV2 model so that the amount of noise can be tuned. To achieve this, we follow the technique originally deployed in Ref.~\cite{Cantori2024}.
Specifically, for each gate, the associated error channel is replaced with a noiseless identity with probability $1-p_{\mathrm{noise}}$. The extreme case $p_{\mathrm{noise}}=0$ corresponds to a completely noiseless circuit, while the other extreme $p_{\mathrm{noise}}=1$ represents the original noise model. The readout error is similarly scaled by a factor $p_{\mathrm{noise}}$, meaning that there is no error when $p_{\mathrm{noise}}=0$, while the error model is unchanged when $p_{\mathrm{noise}}=1$. 
To verify that the tunable FakeMelbourneV2 model is indeed realistic, we make a comparison vis-\'a-vis three actual physical devices available from the IBM Quantum platform. The testbed is a quantum Ising chain with $N=12$ spins. This size is chosen to allow periodic boundary conditions to be implemented on the native topology of the chips. The test quantum circuit is the final stage of a noise-free VQE process. The error in the energy expectation value compared to the ideal case $p_{\mathrm{noise}}=0$ is shown in Fig.~\ref{real}. We find that the three devices perform slightly better than the worst-case scenario $p_{\mathrm{noise}}=1$, reaching performances comparable to those of the regime $0.2\lesssim p_{\mathrm{noise}} \lesssim 0.7$.
As discussed in Sec.~\ref{vqe}, unmitigated VQE would require noise levels reduced by various orders of magnitude compared to this regime.

\subsection{Classically simulable training circuits via partial circuit knitting} \label{knitting_sec}
Circuit knitting techniques provide a suitable strategy for computing the output expectation values of quantum circuits with a large number of qubits~\cite{Mitarai_2021,10236453, Gentinetta2024overheadconstrained}. 
Large circuits are decomposed into smaller, more manageable sub-circuits by removing the gates connecting two or more patches.
This decomposition is particularly useful for near-term quantum devices, where circuit depth and qubit connectivity are limited by hardware constraints, meaning that removing few gates suffices to form disconnected parts.
In our framework, the decomposition leads to small sub-circuits that can be efficiently classically simulated.
In detail, we leverage circuit knitting by cutting two rows of $R_{ZZ}$ gates. 
Taking into account periodic boundary conditions, this partitioning results in two independent sub-circuits, each containing $N/2$ qubits. If necessary, the number of sub-circuits can be increased to reduce their size below the capability of the classical simulator. 
The ideal expectation values of the entire circuit can be reconstructed by combining ({\it knitting}) the results from multiple variations of these smaller sub-circuits.
However, this reconstruction comes at the cost of a potentially large sampling overhead, depending on the number of cut gates and their rotation parameters. Quantitatively, the sampling overhead scales as~\cite{Mitarai_2021}:
\begin{equation}
\label{eqoverhead}
    O = \prod_{\theta_i\in\mathcal{K}} (1+2|\sin(\theta_i)|)^{2} \, ,        
\end{equation}
where $\mathcal{K}$ represents the set of $R_{ZZ}$ gates connecting the two sub-circuits.
This overhead can become prohibitively large, making full circuit reconstruction impractical for deep circuits or large qubit numbers, especially if the angles are sizable. Fortunately, in our framework, it is in fact unnecessary to fully reconstruct the whole circuit. Instead, a subset of $\mathcal{K}$ composed of $C\ge0$ gates can remain unknitted, i.e., their associated parameters $\theta_i$ are set to $0$ or $\pi$. In this way, $C$ cuts are effectively introduced in the original circuit such that, for a sufficiently large $C$, the computation becomes classically feasible. 
This partial circuit knitting approach ensures that the generated training data is representative of the target circuit while controlling the sampling cost. A schematic representation of the knitting procedure is shown in Fig.~\ref{knitting}b. Each circuit is cut at the qubits $q$ and $(q+N/2)\mod N$, and the random index $q$ is uniformly sampled in the interval $q\in[1,N]$. The layers in which the gates are not knitted are, again, chosen randomly. In Sec.~\ref{cost_sec}, we analyse the number of cuts $C$ needed to get accurate results for quantum circuits with different numbers of qubits.

\subsection{Training of the neural networks}\label{training_sec}
We employ standard multilayer perceptrons (MLPs) to predict ideal expectation values from the noisy outputs of the quantum device. For a quantum circuit with $N$ qubits and $P$ layers, we train $2\times N$ independent neural networks to predict $\langle Z_iZ_{i+1}\rangle$ and $\langle X_i\rangle$, for $i=1,\dots,N$. These expectation values allow the computation of the ground-state energy $E$.
The inputs of the MLP network are the noisy expectation values and the sine and cosine of the set of angles $\boldsymbol{\theta} = \{\theta_1, \theta_2, \dots, \theta_{2\times N\times P}\}$ that characterize each quantum circuit.

The training circuits are designed to resemble those encountered in different stages of the VQE optimization.
Starting from the circuit obtained from the previous stage of VQE, random angle variations are applied.
First, since the parameter shift rule is used to determine the gradient, the training set is augmented by including circuits that feature one $\pm \frac{\pi}{2}$ angle shift. Specifically, the shift is applied to one of the parametrized gates, randomly selected with a uniform probability, and with the same probability the circuit without shift is considered.
Moreover, a small random shift is applied to each angle. This is sampled from a normal distribution with zero mean and standard deviation $\sigma=0.05$. 
This additional offset allows using the same neural network for several VQE steps, even when the circuit parameters have sizably changed. The small variance facilitates the training process~\cite{cantori2024challengesopportunitiessupervisedlearning}. 
In our implementation, the training process is repeated every 9 steps of VQE optimization. This number is chosen because, after this many steps, some gate parameters typically shift more than $\sigma$.
It is worth mentioning that transfer learning methods~\cite{Mari2020transferlearningin,Zen_2020} might allow reducing the required amount of training data~\cite{Cantori_2023, PhysRevE.102.033301}.
The training circuits are classically simulated using the circuit knitting technique with two sub-circuits, as discussed in Sec.~\ref{knitting_sec}. 
For large-scale circuits, the sampling technique characteristic of the circuit knitting methods could be adopted. 
However, in our study, 
given the small size of the circuits, we employ a statevector simulation~\cite{quantum_ai_team_and_collaborators_2020_4023103}. 
It is important to emphasize that, in our analysis of the DL-EM method, the computational bottleneck is represented by the classical simulations of the noisy training circuits. In fact, these are computationally more demanding than noise-free simulations. Clearly, when deploying the DL-EM technique on actual quantum devices, these simulations are not needed, allowing scaling to larger qubit numbers.
The statistical noise induced by finite sampling associated with circuit knitting is analysed in Sec.~\ref{cost_sec}.

The training of the MLPs is performed by minimizing the mean-squared-error loss function:
\begin{equation}\label{mse}
	\mathcal{L} = \frac{1}{K_\mathrm{train}}\sum_{k=1}^{K_\mathrm{train}}
 \left(y_k - \tilde{y}_k \right)^2 \, ,
\end{equation}
where $K_{\mathrm{train}}$ is the number of training instances, $y_k$ is the target value, and $\tilde{y}_k$ is the corresponding predicted value.
The network parameters are optimized using the ADAM algorithm. 
To evaluate the prediction accuracy, we compute the coefficient of determination
\begin{equation}\label{r2}
R^2=
1-\frac{
\sum_{k=1}^{K_\mathrm{test}} \left(y_k - \tilde{y}_k \right)^2}{\sum_{k=1}^{K_\mathrm{test}} \left(y_k - \bar{y} \right)^2} ,
\end{equation}
where $\bar{y}$ is the average of the target values and $K_{\mathrm{test}}$ is the number of test instances.
The coefficient of determination quantifies how well the model captures the variations in the target data. 
A normalized error measure is obtained as $1-R^2$,  which corresponds to the ratio of the mean squared error over the data variance.

\section{Results}\label{Sec3}
\begin{figure}
    \centering
    \includegraphics[width=0.8\textwidth]{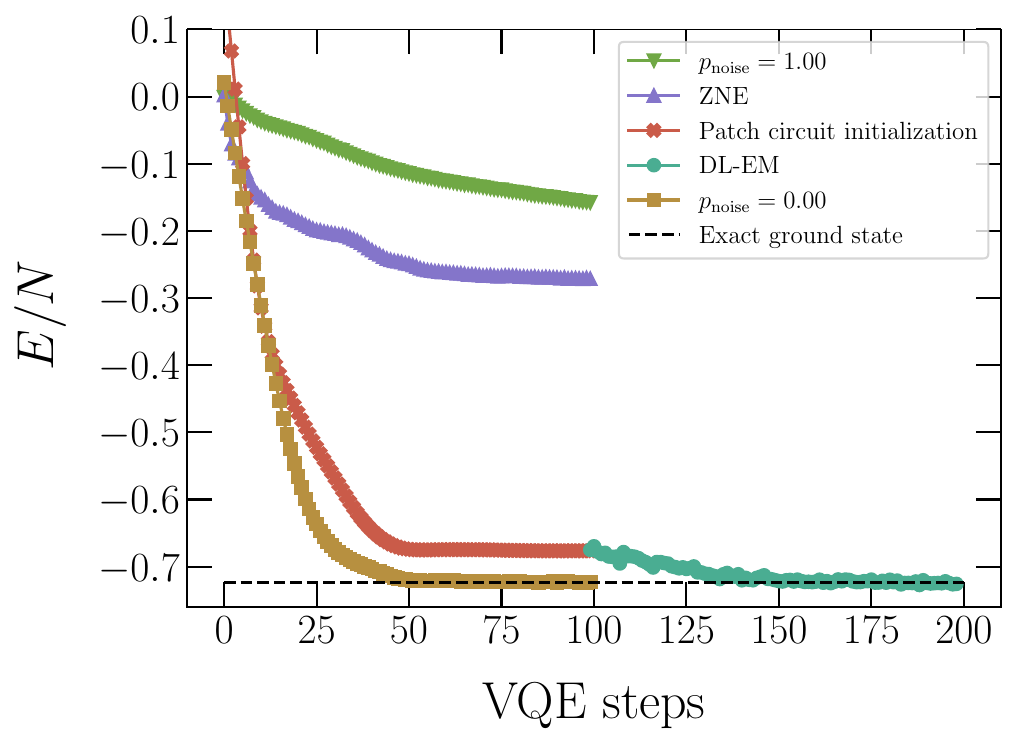}
    \caption{Energy per qubit $E/N$ as a function of the number of VQE optimization steps. We compare the ideal VQE execution ($p_{\mathrm{noise}}=0$), the noisy VQE performed with the FakeMelbourneV2 model ($p_{\mathrm{noise}}=1$), the noisy VQE mitigated using our DL-EM method, and a general-purpose EM method, namely, ZNE. The smart initialization using patch circuits is also shown. The exact ground-state energy is indicated by the horizontal dashed line. 
    VQE is performed using a variational circuit with $P=8$ layers acting on $N=6$ qubits. The DL-EM network is trained on partially knitted circuits with $C=12$ cuts. ZNE is performed using noise scaling factors 1, 3, and 5, corresponding to 0, 1, and 2 repetitions of the quantum circuit and its inverse, with a linear extrapolation method~\cite{mitiq}.}
    \label{vqe}
\end{figure}
\subsection{Testing the VQE algorithm improved by DL-EM}
\begin{figure}
    \centering
    \includegraphics[width=0.8\textwidth]{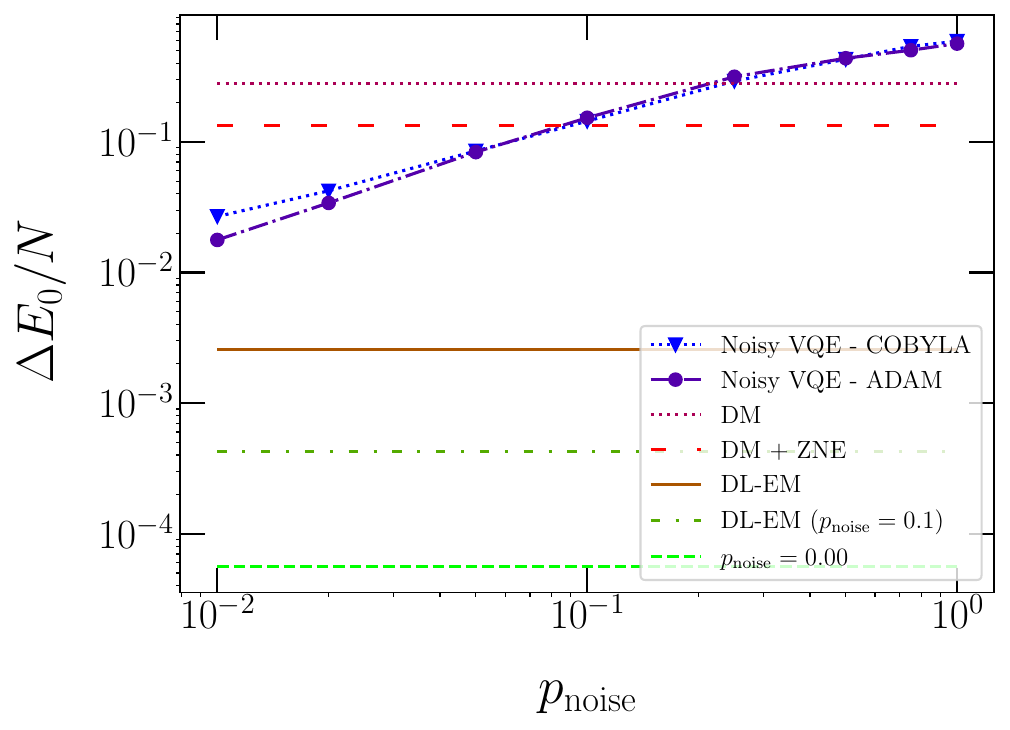}
    \caption{Discrepancy per qubit $\Delta E_0/N$ in the estimate of the ground state energy as a function of the noise level $p_{\mathrm{noise}}$.  The two datasets with symbols correspond to noisy VQE performed with COBYLA and with ADAM optimizers. The horizontal lines denote: DM approaches with and without ZNE, noisy VQE boosted by the DL-EM method executed on the FakeMelbourneV2 model and on a reduced noise model ($p_\mathrm{noise}=0.1$), and the noiseless execution of VQE ($p_\mathrm{noise}=0$). For the DM approaches, the discrepancy is computed applying the correction to the last circuit of VQE performed with the DL-EM method. For the other results, the whole VQE process is performed with the ADAM optimizer, unless otherwise specified.}
    \label{pnoise}
\end{figure}

Hereafter, the ground-state energy of random quantum Ising chains featuring $N=6$ qubits is determined using the VQE algorithm. In particular, the performance of a noisy device model, namely the FakeMelbourneV2 model, is tested using different EM methods. The VQE process for a representative instance of the random Hamiltonian~\eqref{hamiltonian} is visualized in Fig.~\ref{vqe}. 
With the chosen circuit depth, the noiseless execution of VQE rapidly approaches the exact ground state energy $E_0$.
Instead, when performed with the noise model, the optimization remains well above $E_0$. Notably, the DL-EM method restores convergence, as evidenced by the rapid descent after the smart initialization stage. On the other hand, a popular general-purpose EM method, namely, ZNE, provides only a marginal improvement compared to the noisy VQE. 
The final accuracy is better visualized in Fig.~\ref{pnoise}, where the energy discrepancy $\Delta E_0$ is plotted as a function of the noise level $p_{\mathrm{noise}}$.
Even reducing this parameter to $p_{\mathrm{noise}}=0.01$, the energy discrepancy of noisy VQE is still above $1\%$, meaning that VQE performed on near-term quantum devices without EM is impractical for the considered circuit depth.
Gradient-based and gradient-free optimization algorithms provide comparable performances. It is worth mentioning that starting the unmitigated VQE from the patch-circuit initialization, as for the DL-EM approach, doesn't improve the results (not shown).
The DL-EM method reaches an error below the 1\% threshold even when executed with the original noise model, corresponding to $p_{\mathrm{noise}}=1$, and below 0.1\% when the noise level is reduced to $p_{\mathrm{noise}}=0.1$.
Fig.~\ref{pnoise} also analyses the performance of another general-purpose EM strategy, namely, the damping factor method (DM, see Sec.~\ref{method1_sec}), executed with or without ZNE.
To implement this method, we analyse the quantum circuit resulting from the last step of the VQE process performed with the DL-EM.

%
%
To further characterize the performance of the DL-EM method, we repeat the VQE process for four random instances of the quantum Ising chain, at noise level $p_{\mathrm{noise}}=0.1$. This yields an average error of $\Delta E_0/N\simeq9.5\times10^{-4}$, indicating a systematic convergence of VQE with DL-EM.

\subsection{Trade-off between training resources and prediction accuracy in deep learned error mitigation}\label{cost_sec}
\begin{figure}
        \centering
        \includegraphics[width=0.7\linewidth]{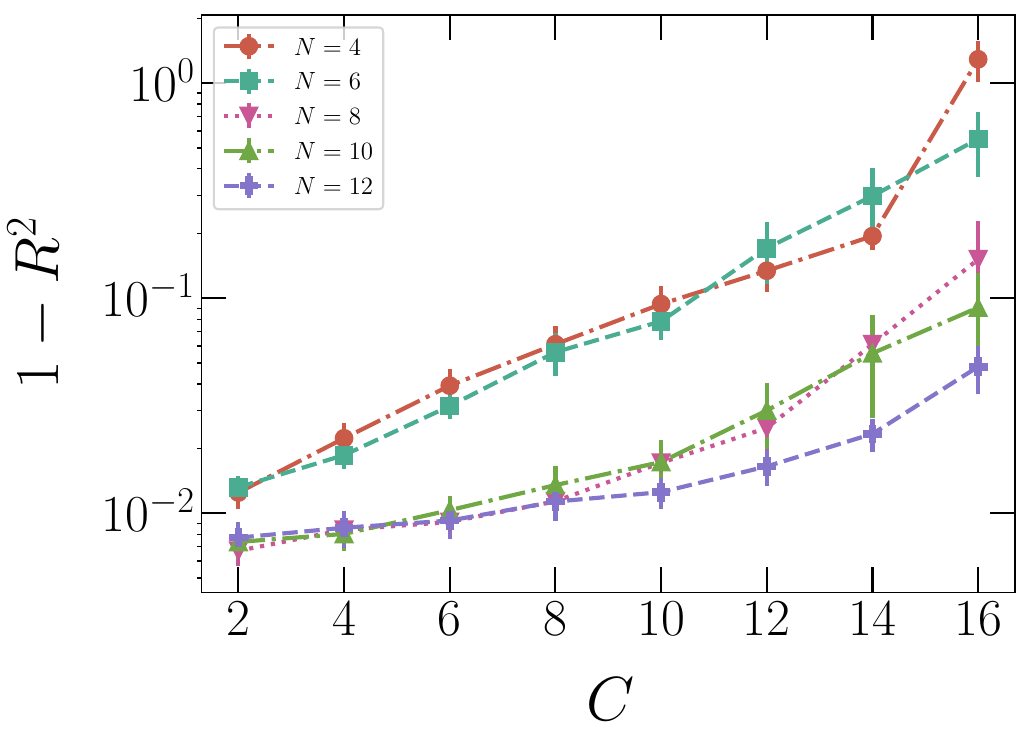}
        \caption{Prediction error $1-R^2$ as a function of the number of cuts $C$ (i.e, the number of not-knitted $R_{ZZ}$ gates) in the training circuits, for different numbers of qubits $N$. 
        The training-set includes $10^4$ circuits, while the test-set includes $10^3$ exactly simulated full quantum circuits, namely, setting $C=0$. 
        }
        \label{cuts}
    \end{figure}

%
%
This section examines how the structural similarity between training and target circuits affects the prediction accuracy. 
This analysis also demonstrates that the classical computational cost for training the DL-EM can be reduced without compromising its accuracy.
One expects that, the more closely the training quantum circuits resemble the target circuit, the better the predictive performance of the model. This expectation is supported by the error measures $1-R^2$  reported in Fig.~\ref{cuts}, as discussed hereafter.
The training and test sets analysed in this section are generated from quantum circuits optimized via simulated VQE processes, each executed for 100 steps with random initialization. Then, offsets and cuts are executed as described in Sec.~\ref{Sec2}. The integer $C$ denotes the number of cuts, i.e., the number of $R_{ZZ}$ gates whose angle parameters $\theta_i$ are fixed to $0$ or $\pi$. Reducing $C$ (i.e., using fewer cuts) results in training circuits that more closely mimic the target circuits, thereby improving prediction accuracy. However, this comes at the cost of increased computational resources during training data generation, due to the larger sampling overhead.
As shown in Fig.~\ref{cuts}, there is a trade-off between the accuracy of the predictions and the computational cost of simulating the training circuits. Notably, the accuracy of DL-EM does not degrade with the size of the target circuit for a fixed $C$.  On the other hand, if the (partial) knitting is implemented with only two patches, the classical simulation cost  scales as $2^{N/2}$.

Some machine learning-based EM methods have already been discussed in the previous literature. In one prominent method, the training is performed on product states~\cite{PhysRevResearch.6.013223}. In our framework, product states are obtained by performing $C=48$ cuts, which is well beyond the regime where the error metric is satisfactory (see Fig.~\ref{cuts}). Likewise, approaches that train using near-Clifford circuits~\cite{Czarnik2021errormitigation,PhysRevResearch.3.033098} are not suitable for our setting. Indeed, these methods require most gates to be Clifford. However, even in the worst-case scenario $C=16$ of Fig.~\ref{cuts}, where the performance is already compromised, 80 of the 96 parameterized gates are non-Clifford, making Clifford-based EM methods inadequate.
Other machine learning-based methods~\cite{Liao2024} that learn how to mimic expectation values mitigated via standard approaches, such as ZNE, are limited by the accuracy of such techniques, and these appear to perform sub-optimally for the analysed circuits.

\begin{figure*}
	\centering
	\includegraphics[width=0.7\textwidth]{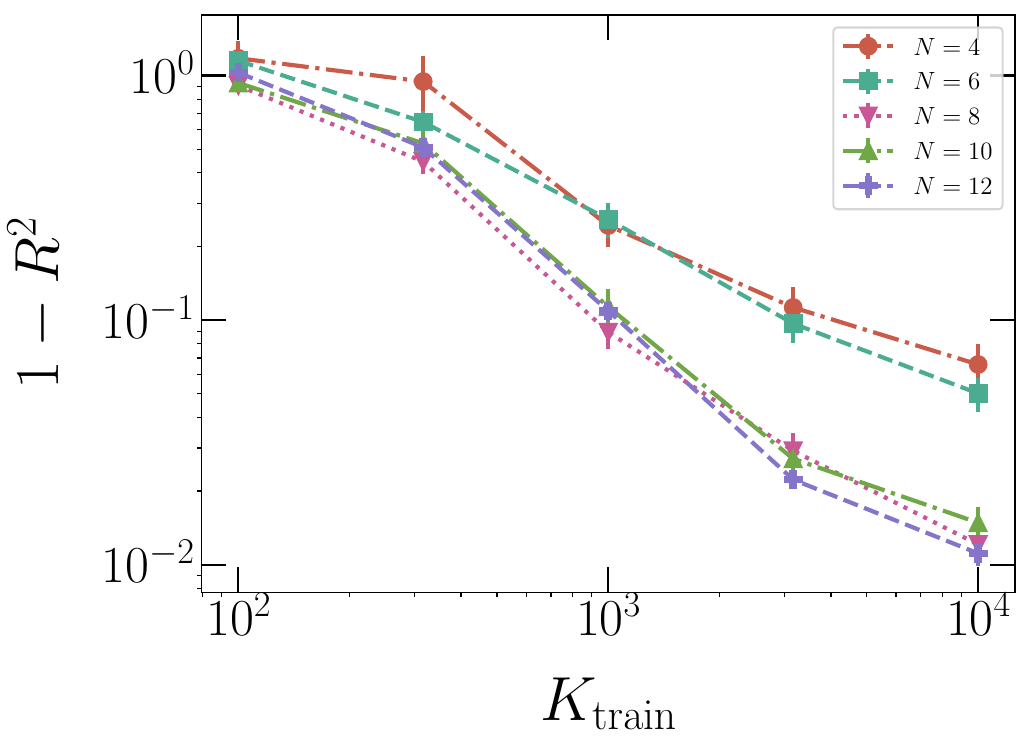}
	\caption{Prediction error $1-R^2$  as a function of the training-set size $K_{\mathrm{train}}$, for different number of qubits $N$. The training circuits feature $C=8$ cuts, while the test-set includes $10^3$ full quantum circuits, namely, setting $C=0$.} 
	\label{Ns}
\end{figure*}

Training deep neural networks requires sufficiently large  datasets, potentially leading to overwhelming computational costs for their generation. In Fig.~\ref{Ns}, we show that the prediction error achieved by our DL-EM model decreases quite rapidly with the training-set size $K_\mathrm{train}$. Importantly, we find that the learning speed does not decrease with the qubit number, suggesting that the DL-EM method could be scaled to larger circuits.

\begin{figure}
        \centering
        \includegraphics[width=0.7\linewidth]{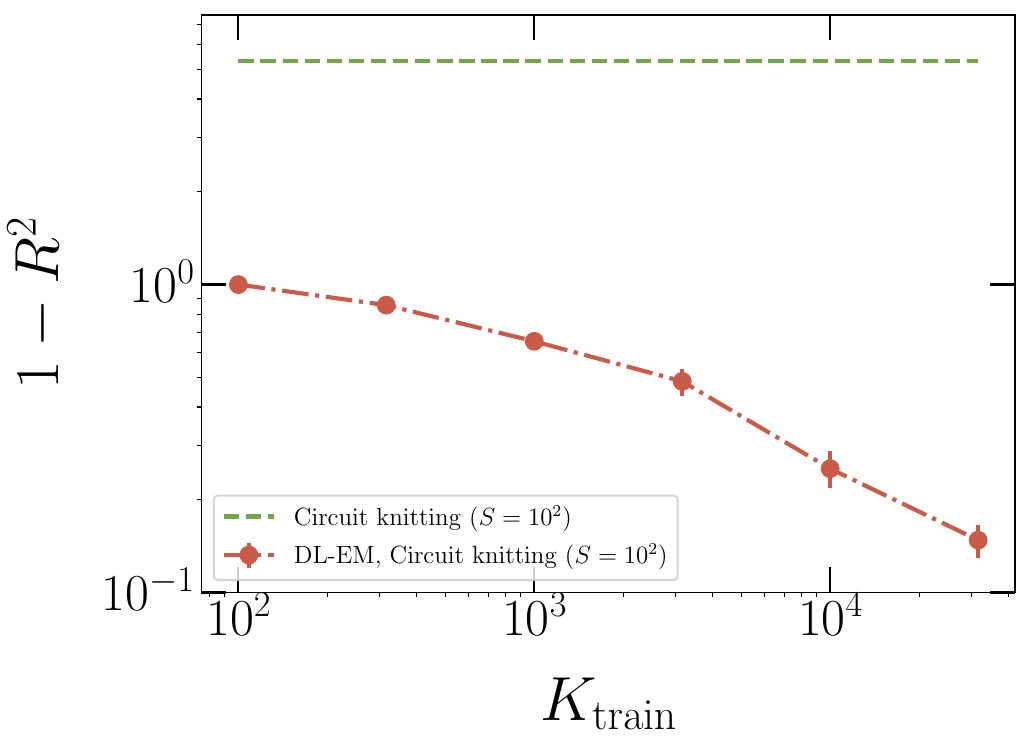}
        \caption{Prediction error $1-R^2$ as a function of the training-set size $K_{\mathrm{train}}$. The DL-EM is trained on target values obtained with the circuit knitting technique with $C=14$ cuts (i.e. only two knitted gates) and $S=10^2$ shots. The model is tested on quantum circuits with the same number of cuts $C=14$, but simulated using exact statevector techniques (i.e. without shot noise). The horizontal line represents the error of the expectation values obtained with the circuit knitting technique with $S=10^2$, measured as the deviation from the statevector results.
        }
        \label{qpd}
    \end{figure}

Finally, it is important to analyse the possible detrimental effect of shot noise. In fact, shot noise is inherent in the use of quantum devices, since expectation values are estimated with a finite number of shots $S$, leading to stochastic fluctuations in these estimates. One might expect that shot noise is particularly impactful in our DL-EM method, since the standard circuit knitting technique in principle requires an exponentially scaling shot number, as predicted by Eq.~\eqref{eqoverhead}. However, it has been found that deep neural networks can often be trained on target values affected by random fluctuations, effectively filtering the noise out~\cite{Pilati_2019, rolnick2018deeplearningrobustmassive}. In particular, this phenomenon was observed in the learning of circuit outputs affected by shot noise~\cite{Cantori_2023}. Indeed, in Fig.~\ref{qpd} we show that the DL-EM model can be trained on target values derived from partial circuit knitting with a small shot number $S$. Although noisy circuit outputs do not accurately approximate the exact expectation values, leading to a large error metric $1-R^2$, the DL-EM model provides more accurate predictions, and these rapidly improve as the size of the training set increases. 

\section{Conclusions}\label{Sec4}
Our work introduces a deep learning-based quantum error mitigation strategy (DL-EM) that significantly improves the performance of variational quantum eigensolvers in the presence of realistic noise. By training deep neural networks on hybrid quantum-classical data, combining noisy expectation values and circuit descriptors, we provide accurate predictions of the ground state energies for the 1D random quantum Ising model. An important aspect is the use of partial circuit knitting, which balances classical computational cost with the need to keep training circuits representative of the target variational circuits, without incurring the exponential overhead of full circuit reconstruction.

Our results demonstrate that DL-EM outperforms general-purpose EM techniques such as ZNE and DM approaches, achieving small energy discrepancies. Notably, the performance further improves as noise is reduced, meaning that DL-EM will strongly benefit from the further development of quantum devices. These findings highlight the critical role of application-specific EM and the synergistic use of quantum resources with classical machine learning.

This work presents DL-EM as a potential approach for improving deep-circuit VQE on near-term quantum hardware by addressing some of the challenges introduced by noise. Future directions include applying the framework to larger systems and more complex Hamiltonians, implementing transfer learning to reduce training costs, and exploring hybrid methods that combine DL-EM with other EM strategies. Further validation across different quantum architectures and systematic scalability studies will be important to assess its practical impact. Overall, these contributions add to the broader effort to improve the reliability and accuracy of variational quantum algorithms on current hardware.

\ack
We acknowledge support from: the MUR PNRR Extended Partnership NQSTI - PE00000023, the MUR PRIN2022 project ``Hybrid algorithms for quantum simulators'' -- 2022H77XB7, 
the National Centre for HPC, Big Data and Quantum Computing (ICSC), CN00000013 Spoke 7 -- Materials \& Molecular Sciences under the project ``INNOVATOR'', and the MUR PRIN-PNRR 2022 project ``UEFA'' -- P2022NMBAJ.
S.P. also acknowledges support from the CINECA awards IsCc2\_REASON and INF25\_lincoln, for the availability of high-performance computing resources and support, and from the EuroHPC Joint Undertaking for awarding this project access to the EuroHPC supercomputer LUMI, hosted by CSC (Finland) and the LUMI consortium, through EuroHPC Development and Regular Access calls.

\bibliographystyle{myunsrturl}
\bibliography{mybibliography}

\end{document}